# ICN-aware Network Slicing Framework for Mobile Data Distribution

Asit Chakraborti, Syed Obaid Amin, Aytac Azgin, Ravishankar Ravindran and Guoqiang Wang
Huawei Research Center, Santa Clara, CA, 95050, USA.
{asit.chakraborti, obaid.amin, aytac.azgin, ravi.ravindran, gq.wang}@huawei.com}@huawei.com

*Abstract*—Network slicing offers an opportunity to realize ICN as a slice in 5G deployment. We demonstrate this through a generic service orchestration framework operating on commodity compute, storage and bandwidth resource pool to realize multiple ICN service slices. Specifically, we show the dynamic creation of real-time audio/video conferencing slices, over which multi-participant communication is enabled. These slices leverage ICN features of name-based routing, integrated security, inherent support for multicasting and mobility, and in-network caching-and-computing to scale and deliver services efficiently, while dynamically adapting to varying service demands. Proposed framework also enables mobility-on-demand feature as a service over an ICN slice to more effectively support producer mobility over multi-access links, such as LTE, Wifi and Ethernet, as will be demonstrated with our demo.

## I. Network Slicing

Network slicing (NS) is proposed in [1] to support diverse service networks in 5G, with services ranging from high bandwidth services (requiring multi-Gbps transmission rates) to low end-to-end latency services (with *1-10ms* latency requirements), and to IoT services scaling over massive number of devices and applications. Slicing entails offering logical and/or physical separation of the service, control and data planes among different services based on virtualized/physical compute-storage-bandwidth resources. This separation is to guarantee isolation in order to satisfy service layer agreements (SLA), which typically include quality of service (QoS), reliability, availability and security. NS follows a service-centric top-down model, which is in contrast to traditional virtual private networks (VPNs) that focus on connectivity isolation in the networking and/or in the lower transport planes. As NS will be based on network softwarization (NSW) at all levels, this will allow dynamic spawning and elastic scaling of service subnetworks based on their corresponding demands. NSW will leverage software defined networking (SDN) and network function virtualization (NFV) frameworks to achieve the goal of programming compute-bandwidth-storage resources among multiple services that span multiple domains. While SDN allows the decoupling of control plane logic from data plane operations to enable arbitrary programmability of the service network, NFV allows the separation of network functions from the underlying hardware implementation[1]. In short, NSW will allow new network functions, hence novel network architectures such as ICN, to co-exist with the existing IP services. Fig. 1 shows a generic network slicing framework that span heterogeneous access networks (wireless and fixed access), edge cloud resources, core transport network and central data centers. This framework provides service orchestration components to support both IP and ICN service slices[2].

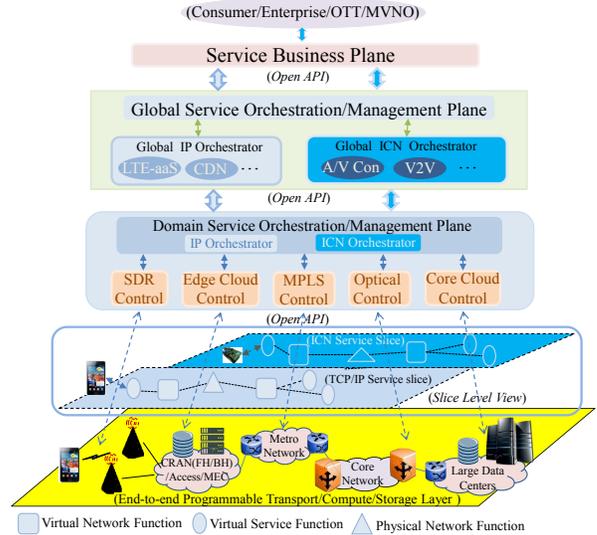

Fig. 1. Network Slicing Framework.

Next, we discuss this framework from the perspective of how ICN slices are realized:

- ICN slice creation begins with service owners providing their SLA requirements to an ICN aware *service business plane* in terms of service quality, reliability, availability and security requirements, which are usually defined using a slice template construct.
- Service requirements are then passed to a corresponding *global ICN service orchestrator*, which converts these requirements to a virtual network service graph, in which the nodes represent ICN service gateways/routers, ICN service functions and storage nodes. These nodes and links are further associated with metadata components such as compute requirements, storage/cache sizes, link QoS metrics (such as bandwidth and latency), security and reliability parameters to support the requirements of the ICN slice.
- Generated service graph is then partitioned into a sub-

---
[1]Generalized processors (such as x-86, GPU and ARM resources) and programmable forwarding plane resources (based on P4/POF) will be used to realize forwarding planes capable of executing different protocol logics.

set of service graphs, referred as subgraphs, and the information is passed to *domain service orchestrators*, which in turn invoke application-agnostic controllers. These controllers offer virtualization of compute-storage-bandwidth resources to convert the service graph requirements to a resource allocation matrix associated with the targeted network components (*i.e.*, access network, edge clouds, inter-connecting transport networks and the regional/central data centers).

The logical to physical layer mapping is technology dependent, and results in provisioning of the necessary resources, which may include ($i$) virtual machines or containers, ($ii$) virtual/physical switch rules, ($iii$) bandwidth resources, ($iv$) security enforcements, and ($v$) queueing management policies, to meet the slice requirements.

Once the ICN service slice is realized, monitoring is enforced dependent on the SLA requirements at each segment of the network, where the slice state exists to ensure smooth functioning of the service(s) over the created slice.

## II. ICN Slicing Demo Architecture

We realized an ICN service slicing prototype following the top-down service creation model discussed earlier. The architecture is shown in Fig. 2, wherein the service plane hosts the ICN service orchestrators, in this case, to manage the audio/video (A/V) conference slices. ICN compute virtualization is handled using Docker Swarm, which manages and orchestrates the containers executing the ICN network and service functions among the available compute resource pool considering ICN slice requirements, while ICN network virtualization is realized over Open Networking Operating System (ONOS). ICN extensions to ONOS include: ($i$) service-specific event-driven routing using *ICN service controllers*, ($ii$) ICN adaptation functionality (*ICN-ONOS adaptor*) to (de-)multiplex messages from these controllers to the hosts, to which the ICN routing instructions are directed, and vice versa. Following are the features we demonstrate through this system:[2]

### A. Feature 1: Audio/Video Slice Orchestration

Realizing each conference in its own slice allows logical separation of PIT/CS/FIB[3] state providing much needed privacy and performance isolation among the conference slices. The A/V conference service architecture is based on [3], which uses in-network service components for synchronizing names among multiple participants. Here, a user requests for a conference slice by providing inputs describing site locations and expected number of participants per site along with parameters that would help in characterizing traffic between sites. This load definition determines the throughput capacity of the virtual forwarders, hence the computing and caching estimates for the slice forwarders. Furthermore, dynamic name-based networking in the ICN forwarders is conducted based on conference service requirements by the conference controller in ONOS, which supports ad hoc join/leave events by participants (each of which is a consumer and/or producer) over a conference slice. Optimization algorithms are also being developed as part of the A/V Service Orchestrator to adapt the infrastructure resources to changing demands of a conference slice.

[2]The video for an earlier version of our demo can be accessed through the following link: https://www.youtube.com/watch?v=Qbq4ps_d_lA.

[3]PIT refers to Pending Interest Table, CS refers to Content Store, and FIB refers to Forwarding Information Base.

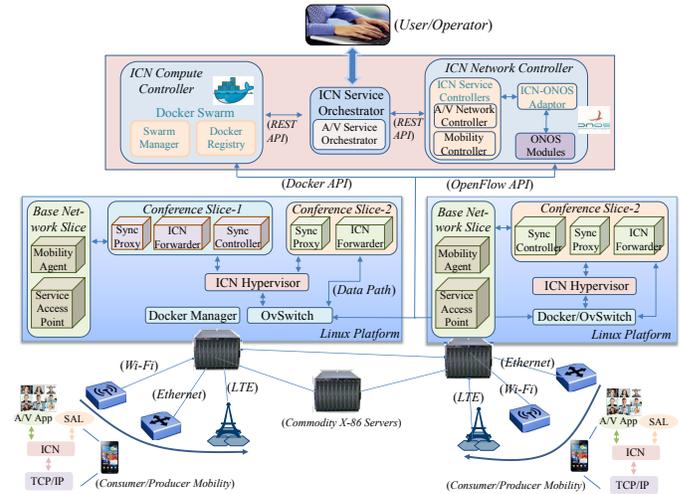

Fig. 2. Demo System Architecture.

### B. Feature 2: Heterogenous Access Mobility

This feature pertains to producer mobility, which is handled based on the ICN network enhancements discussed in [4]. The seamless mobility function is handled by the ICN edge forwarder for the slice (*i.e.*, point of attachment or PoA), which has a new data structure to map name prefixes to topological names for the PoAs. Here, the PoA offers heterogenous access connectivity over LTE (for which the radio implementation uses the open source Open Air Interface or OAI), WiFi and Ethernet links to the participants. PoA handles producer mobility between these heterogeneous access interfaces, while maintaining a user's A/V experience. Consumer mobility is handled by the ICN network through application level re-expression and network caching mechanisms.

### C. Feature 3: Mobility as a Service

Here, we show the co-existence of multiple conference slices, with one over which the mobility service is enabled in a dynamic manner to offer producer mobility service to the participants of that slice. When mobility service is enabled, control plane functionalities allow the PoA to enable late-binding of the Interests towards the new PoA, to support seamless interaction with the mobile producer. This feature also enables updates on the ingress PoA to avoid path stretch.

### D. Feature 4: Slice Management Features

The ICN slice management has multiple views that include: ($i$) the active conference slice view of the slice's virtual network showing the forwarders, service functions and participants; ($ii$) ONOS state view with the ICN forwarder state for each of the slices; and ($iii$) an interface to create/delete conference instances on demand and to turn on/off mobility over a slice.